\documentclass[PRD,twocolumn,showpacs,preprintnumbers,nofootinbib,amsmath,amssymb]{revtex4}

\usepackage{graphicx}
\usepackage{dcolumn}
\usepackage{bm}

\def\fnl{f_{\mathrm{nl}}}
\def\fnle{\widehat \fnl}

\def\VEV#1{\left\langle #1 \right\rangle}

\def\wigner#1#2#3#4#5#6{ \left( \begin{array}{ccc} #1 & #3 & #5
\\ #2 & #4 & #6 \\ \end{array} \right)}

\newcommand{\beq}{\begin{equation}}
\newcommand{\eeq}{\end{equation}}
\newcommand{\beqa}{\begin{eqnarray}}
\newcommand{\eeqa}{\end{eqnarray}}

\begin{document}

\title{The Odd-Parity CMB Bispectrum}

\author{Marc Kamionkowski$^1$ and Tarun Souradeep$^2$}
\affiliation{$^1$California Institute of Technology, Mail Code 350-17,
     Pasadena, CA 91125}
\affiliation{$^2$Inter-University Centre for Astronomy and
     Astrophysics, Pune 411007, India} 

\date{\today}

\begin{abstract}
Measurement of the cosmic microwave background (CMB) bispectrum,
or three-point correlation function, has now become one of the
principle efforts in early-Universe cosmology.  Here we show
that there is a odd-parity component of the CMB bispectrum that
has been hitherto unexplored.  We argue that odd-parity
temperature-polarization bispectra can arise, in principle,
through weak lensing of the CMB by chiral gravitational waves or
through cosmological birefringence, although the
signals will be small even in the best-case scenarios.
Measurement of these bispectra requires only modest
modifications to the usual data-analysis algorithms.  They may
be useful as a consistency test in searches for the usual
bispectrum and to search for surprises in the data.
\end{abstract}


\maketitle

\section{Introduction}

The simplest single-field slow-roll (SFSR) inflationary
models assumed in the now-standard cosmological model predict
departures from Gaussianity to be undetectably small
\cite{localmodel}.  Yet no theorist believes these models to be
the entire story, and many beyond-SFSR models predict departures
from Gaussianity to be larger \cite{larger} and possibly
detectable with current or forthcoming CMB experiments.  Still,
the variety of beyond-SFSR models, and the heterogeneity of
their non-Gaussian predictions, is huge, and no consensus exists
on the likely form of beyond-SFSR physics.  Given the
centrality of this question for physics, however, it is
important to leave no stone unturned; no prospective signal
easily obtainable with existing data, no matter how likely or
unlikely, should be overlooked.

The principle effort in the search for non-Gaussianity is
measurement of the cosmic microwave background (CMB) bispectrum
\cite{bispectrum,Verde},
the three-point correlation function in harmonic space.
Given the small bispectrum signals anticipated, the full
bispectrum is not measured.  Rather, a specific model is
compared against the data to constrain the non-Gaussian
amplitude in that particular model. 
The working-horse model for such analyses has been the
local model \cite{localmodel,Verde}, but the bispectra associated
with a variety other models \cite{others} have also been considered.

The purpose of this paper is to point out that there is an
entirely different class of bispectra that have been hitherto
unexplored.  All bispectrum analyses that have been done so far
assume the bispectrum to be even-parity.  There is, however, an
entirely different class of bispectra that are odd-parity.
Although a odd-parity temperature bispectrum cannot arise from a
projection of a three-dimensional density bispectrum, odd-parity
temperature and temperature-polarization bispectra can arise, at
least in principle, from lensing by gravitational waves or
from cosmological birefringence.  

Although these signals are small---perhaps unobservably
so---they may be worth pursuing for at least two reasons: (1)
The analyses required to determine the standard
(even-parity) bispectrum amplitudes are complicated.  For
example, Ref.~\cite{Yadav:2007yy} claimed evidence for a
non-Gaussian signal in WMAP data, in disagreement with other
null searches \cite{measurements}.  Modifications of the
standard analyses to include measurement of odd-parity bispectra
should be simple and straightforward, and they thus provide,
with the expectation of a vanishing signal, a valuable null test,
and thus consistency check, for the standard searches.  And (2)
there may be new parity-violating physics, beyond what we have
envisioned here, that might give rise to such signals.  It is
with these motivations that we now explore odd-parity bispectra.

To begin, recall that a CMB experiment provides a measurement of
the temperature $T(\hat n)$ as a function of position $\hat n$
on the sky.  The temperature can be re-written in terms of
spherical-harmonic coefficients
$a_{lm} = \int\, d^2 \hat n\, T(\hat n) Y_{lm}^*(\hat n)$.  The
rotationally-invariant CMB power spectrum is $C_l =
\VEV{|a_{lm}|^2}$, where the angle brackets denote an average
over all realizations.  The bispectrum is given by
\begin{equation}
     B_{l_1l_2l_3}^{m_1 m_2 m_3} \equiv \VEV {a_{l_1m_1} a_{l_2m_2}
     a_{l_3m_3}}.
\label{eqn:bispectrum}
\end{equation}
Here we always choose $l_1 \leq l_2 \leq
l_3$, in contrast to most of the literature, which
assumes the bispectrum to be symmetric in $l_1,l_2,l_3$, as
symmetrization wipes out the inherently antisymmetric signals we
consider here.  The rotationally-invariant, or angle-averaged,
bispectrum is
\begin{equation}
     B_{l_1 l_2 l_3} \equiv \sum_{m_1 m_2 m_3}
     \wigner{l_1}{m_1}{l_2}{m_2}{l_3}{m_3} B_{l_1l_2l_3}^{m_1 m_2 m_3} ,
\label{eqn:angleaveraged}
\end{equation}
where the quantities in parentheses are Wigner-3j symbols.
The bispectrum must satisfy the triangle conditions
and selection rules, $m_1+m_2+m_3=0$ and $|l_i-l_j| \leq l_k
\leq |l_i+l_j|$.

The third condition usually assumed in the CMB bispectrum
literature is $l_1+l_2+l_3=$even.  This has nothing to do
with the restrictions of angular-momentum addition encoded in
the Clebsch-Gordan coefficients.  Indeed, one can add, for
example, two angular-momentum states with quantum numbers $l_2=4$ and
$l_3=5$ to obtain a total-angular-momentum state with $l_1=2$.
The restriction $l_1+l_2+l_3=$even is a consequence of the
assumption of parity invariance.  Since the $a_{lm}$ have parity
$(-1)^l$, the bispectrum will have odd parity unless
$l_1+l_2+l_3$=even.

\begin{figure}[htbp]
\centering
\includegraphics[width=0.5\textwidth]{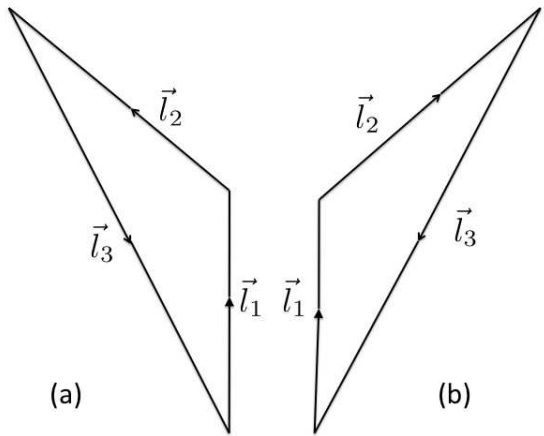}
\caption{Here we plot two Fourier triangles with $l_1<l_2<l_3$
     on a small patch of sky.
     The two have opposite handedness: in (a) the cross
     product $\vec l_1\times \vec l_2$ comes out of the page,
     while in (b) the cross product goes into the page.  The even-parity
     bispectrum (that with $l_1+l_2+l_3=$even) weights both of
     these triangles similarly.  The
     odd-parity bispectrum (configurations with
     $l_1+l_2+l_3$=odd) takes on different signs for the two
     different triangles.}
\label{fig:triangles}
\end{figure}

The distinction between odd- and even-parity configurations can
be understood heuristically for multipole moments $l_i \gg 1$.
On a patch of sky sufficiently small to be approximated as flat,
the three $(l_i,m_i)$ modes then become three plane waves with
wavevectors $\vec l_1,\vec l_2,\vec l_3$.  The conditions
imposed on $(l_i,m_i)$ by the Clebsch-Gordan coefficients then
become a restriction $\vec l_1 + \vec l_2+ \vec l_3=0$.  The
bispectrum then depends on the product of three
Fourier coefficients $T_{\vec l_i}$ for configurations in which
the three wavevectors sum to zero.  Two examples of such
triangles are shown in Fig.~\ref{fig:triangles}, where we have
labeled the triangle sides such that $l_1< l_2< l_3$.
The two triangles are mirror images
of each other. An even-parity bispectrum (that with
$l_1+l_2+l_3=$even) is the same for both of these triangles.  An
odd-parity bispectrum (configurations with $l_1+l_2+l_3$=odd)
takes on different signs for the two different triangles.

Interestingly enough, an odd-parity CMB bispectrum cannot arise
as a projection of a parity-violating density, or potential,
bispectrum, as the distinction between right- and
left-handed triangles does not exist in three spatial
dimensions.  To see this, note that triangle (a) in Fig.~\ref{fig:triangles}
is the same as triangle (b) if we look at it from the other side
of the page.  In other words, in two spatial dimensions, we can
construct a scalar ($\vec l_1 \cdot \vec l_2$) from two vectors
and also a pseudoscalar ($\vec l_1 \times \vec l_2$).  However, in
three spatial dimensions, we can only construct the scalar $\vec
l_1\cdot \vec l_2$ from two vectors.  The
three-dimensional spatial bispectrum therefore has no odd-parity
configurations.  Thus, the condition
$l_1+l_2+l_3=$even on the bispectrum follows simply if we assume
that the CMB map is a projection of a three-dimensional scalar field.

Still, a parity-violating CMB temperature bispectrum might
alternatively arise, for example, if there is a bispectrum for
tensor perturbations (gravitational waves); in this case, the
polarization of one of the the gravitational waves may provide an
additional vector with which to construct parity-violating
correlations.  Lensing by gravitational waves provides a
specific example.  A gravitational wave produces a lensing pattern
that couples two large-$l$ moments $a_{lm}$ due to density
perturbations, but these two are then correlated with the
low-$l$ moment $a_{lm}$ due to the gravitational wave itself
\cite{Dodelson:2010qu}.
This is thus effectively a three-point correlation, and if the
gravitational-wave background is chiral---if there is an
asymmetry in the amplitude of right- versus left-handed
gravitational waves---then the bispectrum may be parity
violating~\cite{Souradeep:2010}.

Other examples can be obtained for three-point
correlations that involve the CMB polarization,
as well as the temperature.
The polarization map is described in terms of spherical-harmonic
coefficients $a_{lm}^E$ and $a_{lm}^B$ for the gradient (E mode)
and curl (B mode) components of the polarization
\cite{Kamionkowski:1996ks},
in addition to the temperature coefficients, which we now call
$a_{lm}^T$.  The parity of the T and E coefficients are
$(-1)^l$, while the parity of the B coefficients are
$(-1)^{l+1}$.  There are now ten three-point correlations that
can be considered (TTT, TTE, TTB, TEE, TBB, TEB, EEE, EEB, EBB,
and BBB), and there are even-parity and odd-parity parts for
each, the parity being determined by $(-1)^{k+\sum_i l_i}$,
where $k$ is the number of B-mode coefficients
\cite{Okamoto:2002ik}.  For example,
$\VEV{a_{l_1m_1}^Ta_{l_2m_2}^Ea_{l_3m_3}^B}$ has even parity for
$l_1+l_2+l_3=$odd and odd parity for $l_1+l_2+l_3=$even.

Suppose that there are no gravitational waves and thus no
B modes at the surface of last scatter.  Density perturbations
will still induce temperature fluctuations and E modes of the
polarization.  If there is a nonzero three-dimensional
bispectrum, for example, of the local-model form, then there
will be even-parity temperature-polarization bispectra induced;
i.e., there will be TTT, TTE, TEE, and EEE bispectra with
$l_1+l_2+l_3$=even.  Now suppose that there is a quintessence field
$\phi$ that couples to the pseudoscalar of electromagnetism
through a Lagrangian term $(\phi/M_*) F \tilde F$, where $F$ and $\tilde F$
are the electromagnetic-field-strength tensor and its dual,
respectively \cite{Carroll:1998zi}.  The time
evolution of $\phi$ then leads to a rotation, by an angle $\alpha =
(\Delta\phi)/M_*$, of the linear polarization of each CMB photon
as it propagates from the surface of last scatter \cite{Carroll:1989vb}.
This rotation then converts some of the E mode into a B mode
\cite{Lue:1998mq}.  If
$\alpha \ll 1$, then these induced B-mode spherical-harmonic
coefficients are $a_{lm}^B \simeq 2\alpha a_{lm}^E$.  This thus
induces, to linear order in $\alpha$, TTB, TEB, and EEB
bispectra with $l_1+l_2+l_3$=even.  But since the parity of the
$a_{lm}^B$ coefficients is opposite to those of the $a_{lm}^T$
or $a_{lm}^E$, the induced TTB, TEB, and EEB bispectra are
parity odd.  Of course, cosmological birefringence will also
induce parity-violating TB and EB power spectra.  Current
constraints \cite{measurements,Feng:2006dp} to the rotation
angle $\alpha$ from these power
spectra, combined with current constraints to the spatial
bispectrum, guarantee that odd-parity bispectra induced by
cosmological birefringence should be small.  

Now that we have discussed some physical mechanisms that might
induce odd-parity bispectra, we now discuss measurement of these
signals.
Implementation of steps in the data analysis to extract these
odd-parity bispectra should be straightforward once the analysis
pipeline for obtaining the even-parity bispectra are in place.
We illustrate with the temperature bispectrum.  It is convenient
to work with the reduced bispectrum, $b_{l_1l_2l_3}\equiv
B_{l_1l_2l_3} /G_{l_1l_2l_3},$
where 
\begin{equation}
     G_{l_1l_2l_3} \equiv
     \sqrt{\frac{(2l_1+1)(2l_2+1)(2l_3+1)}{4\pi}}
     \wigner{l_1}{0}{l_2}{0}{l_3}{0}.
\label{eqn:Gdefn}
\end{equation}
These reduced bispectra, for a given combination of
$l_1+l_2+l_3$=even, can be estimated from the map from
\begin{equation}
   \widehat{b_{l_1l_2l_3}} = G_{l_1l_2l_3}^{-1} \sum_{m_1m_2m_3}
   \wigner{l_1}{m_1}{l_2}{m_2}{l_3}{m_3}
   a_{l_1m_1}a_{l_2m_2}a_{l_3m_3},
\label{eqn:bestimator}
\end{equation}
with variance $\VEV{\left(\widehat{b_{l_1l_2l_3}} \right)^2} =
(G_{l_1l_2l_3})^{-2}$.  

Since measurement of each $b_{l_1l_2l_3}$ will be extremely
noisy, one generally assumes a particular model for the bispectrum and then
estimates the parameter that quantifies the non-Gaussianity.
For example, in the local model \cite{Verde}, $b_{l_1l_2l_3}= 6 \fnl (C_{l_1}
C_{l_2} + \mathrm{perm})$, where $\fnl$ is the non-Gaussianity
parameter.  The minimum-variance estimator for $\fnl$ is then
\begin{eqnarray}
     \fnle &=& \sigma_{\fnl}^2 \sum_{l_1<l_2<l_3} \frac{ 6
     G_{l_1l_2l_3} (C_{l_1}C_{l_2} + \mathrm{perms})}{C_{l_1}^m
     C_{l_2}^m C_{l_3}^m} \nonumber \\
     & & \times \sum_{m_1m_2m_3}
     \wigner{l_1}{m_1}{l_2}{m_2}{l_3}{m_3}
     a_{l_1m_1}a_{l_2m_2}a_{l_3m_3},
\label{eqn:fnle}
\end{eqnarray}
where $C_l^m$ is the power spectrum for the map (including
noise), and
\begin{equation}
      \sigma_{\fnl}^{-2} = \sum_{l_1<l_2<l_3} \frac{ \left[ 6
      G_{l_ll_2l_3} (C_{l_1}C_{l_2} + \mathrm{perms}) \right]^2}{C_{l_1}^m
     C_{l_2}^m C_{l_3}^m},
\label{eqn:fnlevariance}
\end{equation}
is the inverse variance to $\fnle$.  Note that we have
approximated and simplified by restricting $l_1<l_2<l_3$, and
note further that the sums in Eqs.~(\ref{eqn:fnle}) and
(\ref{eqn:fnlevariance}) extend only over multipole moments
$l_1+l_2+l_3$=even.

Measurement of the odd-parity bispectrum is similar, except that we
now sum over configurations with $l_1+l_2+l_3=$odd.  The only subtlety
is that the factors $G_{l_1l_2l_3}$ vanish for $l_1+l_2+l_3=$ odd.  To
remedy this situation, we use identities of Wigner-3j symbols
\cite{varsha} to redefine
\begin{eqnarray}
    && G_{l_1l_2l_3} \equiv {\frac{\sqrt{l_3(l_3+1)
     l_2(l_2+1)}}{\left[l_1(l_1+1)-l_2(l_2+1)-l_3(l_3+1)\right]}}
     \nonumber\\
     &\times&\sqrt{\frac{(2l_1+1)(2l_2+1)(2l_3+1)}{4\pi}}
     \wigner{l_1}{0}{l_2}{-1}{l_3}{1}.
\label{eqn:Gdefnr}
\end{eqnarray} 
This matches Eq.~(\ref{eqn:Gdefn}) for $l_1+l_2+l_3$=even, but remains
non-zero otherwise.  The definition in Eq.~(\ref{eqn:Gdefnr}) is
actually what appears in the bispectrum induced by weak lensing by
chiral gravitational waves~\cite{Souradeep:2010}.

With this replacement, one can then define, for example, an estimator
for an odd-parity bispectrum with a given $l$ dependence (e.g., the
local-model form) through
\begin{eqnarray}
    && \fnle^{\mathrm{odd}} = \sigma_{\fnl}^2 \sum_{l_1<l_2<l_3} \frac{ 6
     G_{l_1l_2l_3} (C_{l_1}C_{l_2} + \mathrm{perms})}{C_{l_1}^m
     C_{l_2}^m C_{l_3}^m} \nonumber \\ 
     &\times &   \sum_{m_1m_2m_3}
     \wigner{l_1}{m_1}{l_2}{m_2}{l_3}{m_3}
     a_{l_1m_1}a_{l_2m_2}a_{l_3m_3}, 
\label{eqn:fnleodd}
\end{eqnarray}
where now the sum is over $l_1+l_2+l_3$=odd.  The variance to
this estimator is again given by Eq.~(\ref{eqn:fnlevariance}),
but now summing over $l_1+l_2+l_3$=odd, and it should be
numerically comparable to the even-parity variance.
Implementation of steps to measure $\fnle^{\mathrm{odd}}$ in an
analysis routine that measures $\fnle$ should be simple and
straightforward.

Some further insight can be gained by considering the form of
estimators for the amplitude of an odd-parity bispectrum in the
flat-sky limit.  We illustrate with a parity-breaking extension
of the local model.  As discussed above, the bispectrum, usually
written as a function $B(l_1,l_2,l_3)$ of the three wavevector
magnitudes, can alternatively be written, taking $l_1<l_2<l_3$,
as a function $B(\vec l_1,\vec l_2)$ of the two shortest
wavevectors.  The usual local model can then be generalized to
\begin{equation}
     B(\vec l_1,\vec l_2) = 2 \left[ \fnl + \fnl^{\mathrm{odd}}
     \frac{ \vec l_1 \times \vec l_2}{l_1l_2} \right] \left
     (C_{l_1} C_{l_2} + \mathrm{perms} \right),
\label{eqn:localmodel}
\end{equation}
where $\fnl^{\mathrm{odd}}$ is an odd-parity non-Gaussian
amplitude.  The minimum-variance estimator for the usual $\fnl$
can then be written in terms of a sum (see, e.g.,
Ref.~\cite{Kamionkowski:2010me}),
\begin{equation}
     \fnle \propto \sum \frac{T_{\vec l_1} T_{\vec l_2} T_{\vec
     l_3} 6(C_{l_1}C_{l_2} +\mathrm{perms})}{C_{l_1}^m C_{l_2}^m
     C_{l_3}^m},
\label{eqn:flatestimator}
\end{equation}
over all triangles $\vec l_1+\vec l_2 +\vec l_3$ with
$l_1<l_2<l_3$.  The minimum-variance estimator for
the odd-parity amplitude $\fnl^{\mathrm{odd}}$ can then be
written analogously as
\begin{equation}
     \widehat{{\fnl^{\mathrm{odd}}}} \propto \sum \frac{T_{\vec
     l_1} T_{\vec l_2} T_{\vec
     l_3} 6(C_{l_1}C_{l_2} +\mathrm{perms})}{C_{l_1}^m C_{l_2}^m
     C_{l_3}^m} \frac{\vec l_1\times \vec l_2}{l_1l_2},
\label{eqn:oddestimator}
\end{equation}
over the same triangles.  In other words, it is the same as the usual
estimator except that it differences, rather than sums, triangles of
different handedness.  Thus, the odd-parity-bispectrum estimator
is a null test for the usual even-parity bispectrum.

To summarize, we have shown that there is a broad class of
odd-parity CMB temperature-polarization bispectra that have been
hitherto overlooked but that can be easily measured with the
data.  We provided two examples of cosmological physics that
could, in principle at least, produce nonvanishing odd-parity
bispectra.  Realistically, though, the bispectra in these
examples will probably be too small to be observed.  Still,
measurement of these odd-parity three-point correlations should
be pursued.  They may provide a valuable
consistency test for the complicated analyses employed to
measure the usual bispectrum amplitude, a null test for
bispectrum measurements analogous to measurements of the curl
\cite{wlcurl} in weak-lensing analyses.  And who knows?  Maybe
there is new parity-violating physics we have not yet foreseen
that might give rise to such signals.  Detection of such a
cosmological signal would, needless to say, be remarkable.

\smallskip
MK thanks the support of the Miller Institute for Basic Research
in Science at the University of California, Berkeley, where part
of this work was completed.  MK was supported at Caltech by DoE
DE-FG03-92-ER40701, NASA NNX10AD04G, and the Gordon and Betty
Moore Foundation.  TS acknowledges support from Swarnajayanti
grant, DST, India and the visit to Caltech during which the work
was initiated.

\end{document}